\documentclass[sigconf]{acmart}

\newif\ifanonymous

\usepackage{xspace}
\usepackage[noabbrev,capitalize]{cleveref}
\usepackage{listings}
\usepackage{fontawesome}
\usepackage[utf8]{inputenc}
\usepackage[T1]{fontenc}
\usepackage{graphicx}
\usepackage{algorithm}
\usepackage{float}
\usepackage{balance}
\usepackage[noend]{algpseudocode}
\usepackage{marvosym}
\usepackage{todonotes}
\usepackage{tikz}
\usetikzlibrary{external}
\usepackage{pgfplots}
\usetikzlibrary{pgfplots.groupplots}
\usetikzlibrary{arrows}
\usetikzlibrary{arrows.meta}
\usetikzlibrary{patterns}
\usetikzlibrary{positioning}
\usetikzlibrary{decorations.pathreplacing}
\usetikzlibrary{shapes.arrows}
\usetikzlibrary{shapes.geometric,shapes.misc}
\usetikzlibrary{pgfplots.groupplots}
\usetikzlibrary{calc}
\usetikzlibrary{fit}
\usetikzlibrary{decorations.pathreplacing,calligraphy}
\pgfplotsset{compat=newest}

\usepackage{titlesec}

\titlespacing\section{0pt}{12pt plus 4pt minus 2pt}{0pt plus 2pt minus 2pt}
\titlespacing\subsection{0pt}{13pt plus 4pt minus 2pt}{0pt plus 2pt minus 2pt}

\usepackage{pifont}
\newcommand{\cmark}{\ding{51}}%
\newcommand{\xmark}{\ding{55}}%

\newfloat{lstfloat}{htbp}{lop}
\floatname{lstfloat}{Listing}
\crefname{lstlisting}{listing}{listings}
\Crefname{lstlisting}{Listing}{Listings}

\usepackage{caption}
\usepackage{subcaption}
\newfloat{listing}{tbhp}{lst}
\floatname{listing}{Listing}
\lstdefinestyle{aarch64}
{
  belowcaptionskip=1\baselineskip,
  breaklines=true,
  numberstyle=\bfseries\color{black},
  language={[x86masm]Assembler},
  showstringspaces=false,
  alsoletter={\#, [,],.},%
  basicstyle=\linespread{0.5}\ttfamily\fontsize{8}{12},
  commentstyle=\itshape\color{gray!70!black},
  identifierstyle=\bfseries\color{black},
  stringstyle=\bfseries\color{black},
  emph={r3,eax,rax,x0,x1,x2,x3,x4,x5,x28,Xptr,Xmod,xzr},
  emphstyle=\bfseries\color{blue},
  classoffset=0,
  morekeywords={adrp,pacia,autia,mov,eor,b,adr,autiza},
  keywordstyle=\bfseries\color{black},
}

\AtBeginDocument{%
  }

\newcommand{\ie}{\textit{i.e.},\ } 
\newcommand{\eg}{e.g.,\ } 
\newcommand{\Eg}{E.g.,\ } 

\usepackage{paralist}
\newcommand{\ReqOne}{\textbf{R}\textbf{1}\xspace}
\newcommand{\ReqTwo}{\textbf{R}\textbf{2}\xspace}
\newcommand{\ReqThree}{\textbf{R}\textbf{3}\xspace}
\newcommand{\ReqFour}{\textbf{R}\textbf{4}\xspace}

\newcommand{\fipac}{\mbox{FIPAC}\xspace}
\newcommand{\fipacos}{\mbox{SFP}\xspace}
\newcommand{\pacia}{\texttt{PACIA}\xspace}

\newcommand{\specgeoruntime}{\mbox{18.8\,\%}\xspace}
\newcommand{\specgeoruntimefipacos}{\mbox{20.6\,\%}\xspace}
\newcommand{\microoverhead}{\mbox{1.9\,\%}\xspace}
\newcommand{\specdiff}{\mbox{1.8\,\%}\xspace}

\copyrightyear{2022}
\acmYear{2022}
\setcopyright{rightsretained}
\acmConference[HASP '22]{Hardware and Architectural Support for
Security and Privacy}{October 1, 2022}{Chicago, IL, USA}
\acmBooktitle{Hardware and Architectural Support for Security and
Privacy (HASP '22), October 1, 2022, Chicago, IL, USA}
\acmDOI{10.1145/3569562.3569565}
\acmISBN{978-1-4503-9871-8/22/10}

\begin{document}
\title[\fipacos: System Call Flow Protection]{\fipacos: Providing System Call Flow Protection against Software and Fault Attacks}

\ifanonymous
  \author{Anonymous Author(s)}
  \affiliation{
    \institution{}
    \country{}
  }
\else
  \author{Robert Schilling}
  \email{robert.schilling@iaik.tugraz.at}
  \affiliation{%
    \institution{Graz University of Technology}
    \city{Graz}
    \country{Austria}
  }

  \author{Pascal Nasahl}
  \email{pascal.nasahl@iaik.tugraz.at}
  \affiliation{%
    \institution{Graz University of Technology}
    \city{Graz}
    \country{Austria}
  }

  \author{Martin Unterguggenberger}
  \email{martin.unterguggenberger@lamarr.at}
  \affiliation{%
  \institution{Graz University of Technology}
  \institution{Lamarr Security Research}
    \city{Graz}
    \country{Austria}
  }

  \author{Stefan Mangard}
  \email{stefan.mangard@iaik.tugraz.at}
  \affiliation{%
    \institution{Graz University of Technology}
    \institution{Lamarr Security Research}
    \city{Graz}
    \country{Austria}
  }

  \renewcommand{\shortauthors}{Schilling et al.}
\fi

\begin{abstract}
  With the improvements in computing technologies, edge devices in the Internet-of-Things or the automotive area have become more complex.
  The enabler technology for these complex systems are powerful application core processors with operating system support, such as Linux, replacing simpler bare-metal systems.
  While the isolation of applications through the operating system increases the security, the interface to the kernel poses a new threat.
  Different attack vectors, including fault attacks and memory vulnerabilities, exploit the kernel interface to escalate privileges and take over the system.

  In this work, we present \fipacos, a mechanism to protect the execution of system calls against software and fault attacks providing integrity to user-kernel transitions.
  \fipacos provides system call flow integrity by a two-step linking approach, which links the system call and its origin to the state of control-flow integrity.
  A second linking step within the kernel ensures that the right system call is executed in the kernel.
  Combining both linking steps ensures that only the correct system call is executed at the right location in the program and cannot be skipped.
  Furthermore, \fipacos provides dynamic CFI instrumentation and a new CFI checking policy at the edge of the kernel to verify the control-flow state of user programs before entering the kernel.
  We integrated \fipacos into \fipac, a CFI protection scheme exploiting ARM pointer authentication.
  Our prototype is based on a custom LLVM-based toolchain with an instrumented runtime library combined with a custom Linux kernel to protect system calls.
  The evaluation of micro- and macrobenchmarks based on SPEC 2017 show an average runtime overhead of \microoverhead and \specgeoruntimefipacos, which is only an increase of \specdiff over plain control-flow protection.
  This small impact on the performance shows the efficiency of \fipacos for protecting all system calls and providing integrity for the user-kernel transitions.
\end{abstract}

\ccsdesc[500]{Security and privacy~Tamper-proof and tamper-resistant designs}
\ccsdesc[500]{Security and privacy~Side-channel analysis and countermeasures}

\keywords{System Call Flow Protection, Control-Flow Integrity, Fault Attacks.}

\maketitle

\section{Introduction}
\label{sec:fipacos:introduction}
Devices in the Internet-of-Things, automotive area, or industrial computers are getting more complex and powerful.
While in the past, those systems used deeply embedded processing units with bare-metal applications, they now are based on powerful application-grade processors with the support for operating systems~\cite{qualcommiot}.
Off-the-shelf operating systems, \eg Linux, build the foundation for complex software~\cite{ubuntucore}.
They isolate different programs, manage privileges, or restrict access to particular memory regions.
User programs can only access kernel features via a small but well-defined interface, the system call~(syscall) interface.
For this reason, this interface to the kernel is a prominent target for attackers to escalate privileges and gain access to the system~\cite{DBLP:conf/fdtc/TimmersM17}.

One way of manipulating the system call interface is control-flow hijacking, which can be conducted with different methodologies.
Classical control-flow attacks performed in software exploit a memory vulnerability to modify a code-pointer or return address on the stack to redirect the execution of the program.
When fault attacks are considered in the threat model, the attack surface in the kernel interface increases even more.
While faults can manipulate the control-flow on a much finer granularity, \eg they can manipulate direct branches, they can also manipulate the system call being executed.
A control flow hijack can skip or change which system call gets executed, possibly with a critical security impact.
Furthermore, precise faults can directly manipulate which system call gets executed by manipulating the system call register containing the system call number.

One way to counteract control-flow attacks is a generic mechanism called control-flow integrity~(CFI)~\cite{DBLP:conf/ccs/AbadiBEL05}.
CFI exists at different granularities, depending on which threat model is considered.
In a classical software setting, only indirect branches are protected since those are the only ones an attacker can manipulate.
Faults pose a more severe threat, thus requiring even more robust protection.
Fine-grained instrumentation~\cite{DBLP:journals/tr/OhSM02a,DBLP:journals/tpds/AlkhalifaNKA99, DBLP:conf/dft/GoloubevaRRV03} protects the control-flow of a program on basic-block or even instruction-level~\cite{DBLP:conf/eurosp/WernerUSM18,DBLP:journals/compsec/ClercqGUMV17}.
As a result, these countermeasures protect direct or indirect branches or even the whole instruction sequence.
Instruction-granular protection requires intrusive hardware changes to deal with the performance penalty, which is unsuitable for commodity devices.

CFI can be enforced in different security domains.
While traditionally, CFI was only used to protect user-space applications, different CFI protection schemes can also protect the kernel~\cite{DBLP:conf/eurosp/GeTPJ16, DBLP:conf/sp/CriswellDA14}.
However, currently, there are no CFI protection schemes available providing protection between different security domains, \ie the transitions between the user-space program and the kernel.
Thus, the large attack surface, the transitions between user programs to the kernel remain unprotected.
Hence, there is a need for new countermeasures that protect the software interface to the kernel and provides system call flow integrity for commodity devices.
\vspace{-2mm}
\subsection*{Contribution}

In this work, we solve the problem of the unprotected system call interface and provide system call flow protection on top of CFI, protecting the interface to the kernel against both software and fault attacks.
\fipacos cryptographically links the system call itself and its origin to a global CFI state that is verified at runtime in the operating system.
A second-stage linking mechanism within the kernel dynamically applies a second link to ensure that the correct system call was selected and executed.

To automatically protect arbitrary programs, we develop an LLVM-based toolchain to provide CFI and instrument all system calls.
We provide an instrumented standard library, where all system calls are instrumented with our system call protection.
Furthermore, we modify the Linux kernel to dynamically verify at runtime that the correct system call was executed.

We implement \fipacos on top of \fipac, a software-based CFI scheme exploiting ARM pointer authentication.
We evaluate the performance of \fipacos based on a microbenchmark to measure the impact of \fipacos on the system call latency, leading to an overhead of \microoverhead.
To show the applicability to real-world programs, we perform macrobenchmarks using the SPEC 2017 application benchmark.
On average, we measure a runtime overhead of \specgeoruntimefipacos for protected applications.
Summarized, we make the following contributions:

\begin{itemize}
	\item We provide system call flow protection by linking the syscall and its origin to a global CFI state and verifying it at runtime.
	\item We provide a prototype implementation comprising an LLVM-based toolchain, an instrumented C-standard library, and a modified Linux kernel.
	\item We evaluate the performance based on a microbenchmark and on the application-grade SPEC~2017 benchmark.
\end{itemize}


\vspace{-3mm}
\section{Background}
\label{sec:fipacos:background}

This section provides background to fault attacks, pointer authentication, and control-flow integrity.
\vspace{-3mm}
\subsection{Fault Attacks}
Injecting faults into a digital circuit is a powerful threat allowing adversaries to break the security of a system entirely.
The effect of an induced fault at the electrical level includes timing violations and transient voltage and current changes~\cite{DBLP:journals/iacr/Richter-Brockmann21}.
Typically, the effect of a fault is modeled at the bit-level with transient bit-flips and permanent stuck-at effects~\cite{DBLP:conf/fdtc/VerbauwhedeKS11}.

Common fault injection approaches include voltage or clock glitching, laser fault injection~(LFI), and electromagnetic fault injection~(EMFI)~\cite{DBLP:journals/iajit/ZiadeAV04}.
While these methodologies require physical access to the device, recently, new techniques relaxing this constraint have been released~\cite{DBLP:conf/uss/TangSS17, DBLP:conf/sp/MurdockOGBGP20, DBLP:conf/uss/ChenVMDOG21,DBLP:conf/ccs/QiuWLQ19}.
\Eg in Plundervolt~\cite{DBLP:conf/sp/MurdockOGBGP20}, the attacker utilizes the dynamic voltage scaling interface of the CPU to induce faults remotely in software.

Independently of the injection technique, an attacker can exploit the effects of faults in various ways.
\Eg fault attacks on encryption primitives enable the attacker to leak secret keys~\cite{DBLP:conf/crypto/BihamS97, DBLP:conf/acisp/ChenY03, DBLP:journals/tches/DobraunigEKMMP18}.
Despite dedicated attacks on encryption, fault attacks are also actively used to bypass security features, such as secure boot, on embedded systems~\cite{timmers2016bypassing, DBLP:conf/woot/CuiH17, pareja2018fault, DBLP:journals/iacr/OFlynn20, DBLP:journals/tches/HerrewegenOGT21}.
By inducing targeted faults into the program counter of a processor, faults enable an adversary to arbitrarily hijack the control-flow of a program~\cite{DBLP:conf/fdtc/TimmersSW16,DBLP:conf/fdtc/TimmersM17, nasahl2019attacking}.
\vspace{-3mm}
\subsection{Control-Flow Integrity}

The control-flow of a program can be hijacked using software attacks, fault attacks, or combined software-fault attacks.
Therefore, various countermeasures targeting different attacker models were proposed to protect programs from these attack vectors.

\vspace{-1mm}
\paragraph*{Software CFI Schemes.}
Software-based control-flow attacks are typically performed by exploiting a memory vulnerability.
By overwriting control-flow-related data, \eg return addresses or function pointers, the adversary can arbitrarily manipulate the execution of the program~\cite{DBLP:conf/ccs/Shacham07, DBLP:conf/ccs/BletschJFL11, DBLP:conf/sp/HuSACSL16}.
To mitigate these attacks, software control-flow integrity~(SCFI) schemes~\cite{DBLP:conf/uss/CowanBJW03, DBLP:conf/osdi/KuznetsovSPCSS14, DBLP:conf/ccs/MashtizadehBBM15, DBLP:conf/uss/LiljestrandNWPE19} aim to provide pointer integrity using different mechanisms.
\Eg PARTS~\cite{DBLP:conf/uss/LiljestrandNWPE19} uses ARM pointer authentication~(PA) to cryptographically seal and verify security-sensitive pointers to protect them while stored in memory.

\vspace{-1mm}
\paragraph*{Fault CFI Schemes.}
Software CFI schemes only protect control-flow transfers the adversary can also manipulate in the software threat model, \ie return addresses and function pointers.
Faults also allow the attacker to tamper with static control-flow data stored in the program or even skip instructions.
Therefore, fault control-flow integrity~(FCFI) schemes enforce their protection at a finer granularity, \eg at the instruction level~\cite{DBLP:journals/compsec/ClercqGUMV17, DBLP:conf/eurosp/WernerUSM18}.
However, as these schemes usually require custom hardware changes to avoid tremendous runtime overheads, software-based FCFI schemes typically operate at the function or basic block level~\cite{DBLP:journals/tr/OhSM02a, DBLP:conf/cgo/ReisCVRA05}.
These schemes track the execution of the program using a signature and compare this running signature with a precomputed signature during runtime.

\vspace{-1mm}
\paragraph*{Software Fault CFI Schemes.}
As most FCFI schemes~\cite{DBLP:journals/tr/OhSM02a, DBLP:conf/cgo/ReisCVRA05} do not consider a software attacker in their threat model, software attacks allow the adversary to bypass most FCFI schemes.
Here, the adversary uses a memory bug to overwrite the state maintained in software and arbitrarily hijacks the control-flow.
Hence, mitigating software, fault, and software-fault combined attacks require even stronger countermeasures, \ie software fault CFI~(SFCFI) schemes.
\vspace{-3mm}
\subsection{\fipac}
\fipac~\cite{DBLP:conf/cosade/SchillingNM22} is a software fault CFI~(SFCFI) scheme mitigating software and fault-based control-flow attacks exploiting ARM pointer authentication.
Internally, \fipac maintains a global state through the entire program execution.
When entering a basic block, \ie a block of consecutive instructions without a control-flow transfer, \fipac cryptographically updates the state.
Depending on \fipac's configured checking policy, the value of the state is compared to the expected value determined during the compilation of the program at the end of each basic block, function, or program.
On control-flow merges, \ie indirect calls, the state is updated using a justification signature to ensure that different valid control-flow paths yield an identical state.
To prevent a software adversary from predicting and overwriting the state using a memory bug, a MAC is utilized for the state update.
Moreover, the state update and check functions cryptographically derive and verify the running signature on program execution.
\fipac uses the pointer authentication instructions of modern ARMv8.6A architectures for the MAC computation.
\vspace{-1mm}
\paragraph*{ARM Pointer Authentication.}
ARM pointer authentication is a hardware feature introduced with ARMv8.3A~\cite{armv83} and updated in ARMv8.6A~\cite{armv86}.
This extension provides new instructions to cryptographically sign and authenticate data.
These instructions derive a message authentication code~(MAC) using a secret key, a 64-bit modifier, and the value of a provided register, \eg an address stored in a pointer.
A fraction of this MAC, called the pointer authentication code~(PAC), is then stored in the upper bits of the provided register.
By using the authentication instructions, the authenticity of the MAC and the data in the register can then be verified.
\vspace{-3mm}
\subsection{Linux and the System Call Interface}

Linux~\cite{linux} is a monolithic kernel used in billions of devices~\cite{linux_usage} and embedded systems.
To retrieve a particular service or get a specific resource, \eg reading and writing a file, or to get dynamic memory, the user program needs to request this from the kernel, \ie via a system call.
A system call changes the privilege and transfers the execution from the user-space program to the kernel of the operating system, which then grants or denies the requested service.
A user-space program aiming to execute a certain system call invokes the corresponding system call wrapper routine provided by a library.
This wrapper then initiates a control-flow and privilege transfer into the kernel space by using a dedicated instruction, \ie the \texttt{svc} instruction for AArch64.
The system call instruction requires the system call number of the requested service and additional optional parameters as arguments.


\vspace{-3mm}
\section{Threat Model and Attack Scenario}
\label{sec:fipacos:threat}

Our threat model considers a powerful adversary capable of performing software attacks, fault attacks, or combined software and fault attacks.
This attacker can exploit memory vulnerabilities to arbitrarily read or modify data in memory.
However, we assume that the code segment of the program cannot be modified by a software adversary by, for example, exploiting memory vulnerabilities.
Nevertheless, by inducing faults, the attacker can flip bits in memory, the registers, the code segment, or the instruction pipeline of the processor.
We assume that the control-flow of executed programs \textit{and} the kernel is protected using an SFCFI scheme, such as \fipac.

Note that faults on the data, except the syscall register, are out of the scope of this work.
It requires orthogonal schemes, \eg redundancy encoding schemes for data~\cite{DBLP:journals/tc/Brown60}, for their protection.
We assume ARM PA to be cryptographically secure, and the attacker does not have access to the encryption keys.
Furthermore, the operating system is assumed to be secure, providing isolation of the kernel task structure to the user program.
\vspace{-3mm}
\subsection{Attack Scenario}

\begin{figure}[t]
  \centering
  \begingroup
  \tikzset{every picture/.style={scale=0.75}}%
  \definecolor{fancygreen}{HTML}{78b473}
\definecolor{fancyred}{HTML}{f70146}
\definecolor{fancygray}{HTML}{a5a5a5}

\newcommand\frameX{1}
\newcommand\frameY{-0.5}
\newcommand\frameLength{12}
\newcommand\frameHeight{4}

\newcommand\syscallX{0.1+1.25}
\newcommand\syscallY{4}
\newcommand\syscallLength{3}
\newcommand\syscallHeight{1}

\newcommand\syscallGap{0.3}

\newcommand\userX{0.1+1.25+1.5-1}
\newcommand\userY{-1}
\newcommand\userLength{2}
\newcommand\userHeight{3}

\newcommand\ptrXone{0.1+3.5+1-0.5-0.25}
\newcommand\ptrYone{1-1}

\newcommand\ptrXtwo{0.1}
\newcommand\ptrYtwo{3}

\begin{tikzpicture}[scale=1.0]
    \tikzset{edge/.style=->, >=stealth, thick};

    \draw[rounded corners=6pt, fill=fancygray!30] (2.2,0.5) rectangle node[scale=0.6]{} ++(1.8,2.0);
    \draw[rounded corners=6pt, draw=none] (2,2.5) rectangle node[scale=0.8]{\textbf{User mode}} ++(4.5,0.5);
    \draw[rounded corners=6pt, densely dotted] (2,0) rectangle node[scale=0.6]{} ++(4.5,3);

    \draw[rounded corners=6pt, draw=none] (2.2,0) rectangle node[scale=0.6]{Application} ++(1.8,0.5);
    \draw[rounded corners=2pt, draw=none] (2.2,1.6) rectangle node[scale=0.6]{\texttt{...}} ++(1.8,0.3);
    \draw[rounded corners=2pt, draw=none] (2.2,1.3) rectangle node[scale=0.6]{\texttt{syscall\_C()}} ++(1.8,0.3);
    \draw[rounded corners=2pt, draw=none] (2.2,1.0) rectangle node[scale=0.6]{\texttt{...}} ++(1.8,0.3);

    \draw[rounded corners=6pt, draw=none] (4.5,0) rectangle node[scale=0.6]{\textit{libc} syscall wrapper} ++(1.8,0.5);
    \draw[rounded corners=6pt, fill=fancygray!30] (4.5,0.5) rectangle node[scale=0.6]{} ++(1.8,2.0);
    \draw[rounded corners=2pt, draw=none] (4.5,1.9) rectangle node[scale=0.5]{\texttt{syscall\_C() \phantom{a}\{}} ++(1.8,0.3);

    \draw[rounded corners=2pt, draw=none] (4.5,1.7) rectangle node[scale=0.5]{\texttt{...}} ++(1.8,0.2);
    \draw[rounded corners=2pt, draw=none] (4.5,1.5) rectangle node[scale=0.5]{\texttt{syscall args}} ++(1.8,0.2);
    \draw[rounded corners=2pt, draw=none] (4.5,1.3) rectangle node[scale=0.5]{\texttt{syscall numb}} ++(1.8,0.2);
    \draw[rounded corners=2pt, draw=none] (4.5,1.1) rectangle node[scale=0.5]{\texttt{svc}} ++(1.8,0.2);
    \draw[rounded corners=2pt, draw=none] (4.5,0.9) rectangle node[scale=0.5]{\texttt{...}} ++(1.8,0.2);

    \draw[rounded corners=2pt, draw=none] (4.5,0.7) rectangle node[scale=0.6,left]{\texttt{\}\phantom{aaaaa}}} ++(1.8,0.3);


    \draw[rounded corners=6pt, draw=none] (7.5,2.5) rectangle node[scale=0.8]{\textbf{Kernel mode}} ++(4.5,0.5);
    \draw[rounded corners=6pt, densely dotted] (7.5,0) rectangle node[scale=0.6]{} ++(4.5,3);

    \draw[rounded corners=6pt, draw=none] (7.7,0) rectangle node[scale=0.6]{Syscall handler} ++(4.1,0.5);
    \draw[rounded corners=6pt, fill=fancygray!30] (7.7+0.5,0.5) rectangle node[scale=0.5]{} ++(3.1,2.0);

    \draw[rounded corners=2pt, draw=none] (7.7+0.5,1.9) rectangle node[scale=0.5]{\texttt{syscall\_C service routine}} ++(3.1,0.3);
    \draw[rounded corners=2pt, draw=none] (7.7+0.5,1.6) rectangle node[scale=0.5]{\texttt{...}} ++(3.1,0.3);
    \draw[rounded corners=2pt] (7.7+0.5,1.6) rectangle node[scale=0.6]{} ++(3.1,0.6);

    \draw[rounded corners=2pt, draw=none] (7.7+0.5,1.3) rectangle node[scale=0.5]{\texttt{syscall\_B service routine}} ++(3.1,0.3);
    \draw[rounded corners=2pt, draw=none] (7.7+0.5,1.0) rectangle node[scale=0.5]{\texttt{...}} ++(3.1,0.3);
    \draw[rounded corners=2pt, draw=none] (7.7+0.5,0.7) rectangle node[scale=0.5]{\texttt{sys\_exit}} ++(3.1,0.3);
    \draw[rounded corners=2pt] (7.7+0.5,0.7) rectangle node[scale=0.6]{} ++(3.1,0.9);


    \draw[edge] (4,1.5) -- ++(0.5,0.6);
    \draw[edge, dotted] (6.3,1.3) -- ++(1.4+0.5,0.8);
    \draw[edge] (6.3,1.3) -- ++(1.4+0.5,0.2);

    \draw[edge] (7.7+0.5,0.9) -- ++(-1.4-0.5,0.3);
    \draw[edge] (4.5,0.9) -- ++(-0.5,0.45);

    \node[orange] at (7,1.7) {\Huge\Lightning};

\end{tikzpicture}%
  \endgroup
  \setlength{\abovecaptionskip}{1mm}
  \setlength{\belowcaptionskip}{-3mm}
  \caption{Redirecting a system call using fault attacks.}
  \label{fig:fipacos:threat:syscall_threat}
\end{figure}
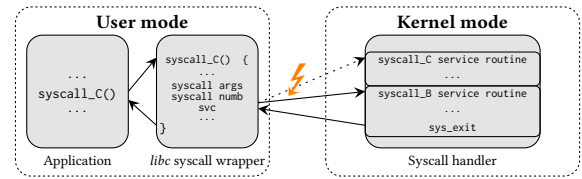

Within this threat model, the adversary aims to hijack the program's interface to the Linux kernel.
In the example shown in \Cref{fig:fipacos:threat:syscall_threat}, the user program invokes the system call \textbf{C} using the Linux system call interface.
However, by using a fault attack or a software-fault combined attack, the adversary can either \textit{(i)} redirect the system call to \textbf{B} or \textit{(ii)} entirely skip the system call.

\begin{lstlisting}[float,language=c++,style=aarch64,caption={Invoking system call \textbf{C} on AArch64.},label={lst:fipacos:writesyscall}, frame=bt,aboveskip=0mm,belowskip=-5mm]
basic_block:
  ...
  ldr w8, memAddress  ; load data from memAddress to w8
  ...
  mov x0, #...       ; arguments for C system call
  mov w8, #syscall_C  ; system call number for C
  svc #0
\end{lstlisting}
\Cref{lst:fipacos:writesyscall} shows the instruction sequence to invoke the system call \textbf{C} on AArch64.
The system call number is stored in register \texttt{w8}, and the system call arguments are stored in the remaining registers.
By flipping bits in register \texttt{w8} using faults, the adversary can redirect \textit{(i)} the execution to a different system call.

Moreover, the syscall gadget in \Cref{lst:fipacos:writesyscall} is susceptible to combined attacks.
A software-fault combined attacker utilizes a memory vulnerability to overwrite data at address \texttt{memAddress}.
Afterward, in Line~4, the adversary hijacks the execution of the program by flipping bits in the program counter to redirect the control-flow to the \texttt{svc} instruction in Line~7, responsible for switching to the kernel.
This attack enables the adversary to invoke arbitrary system calls.
In addition to these attacks, a fault attacker can also corrupt the \texttt{svc} instruction to skip \textit{(ii)} the execution of the entire syscall.

SCFI schemes, such as \fipac, currently \textit{cannot} mitigate these attacks as these countermeasures do not consider transitions between user-space and kernel space in their threat model.
While they only protect the user-space application, they fail to provide protection for the kernel interface, posing a large threat surface for critical vulnerabilities.
Furthermore, current SCFI protection schemes use static control-flow instrumentation, which is the same for subsequent calls to the program.
As a result, an attacker with access to the code segment or to general-purpose registers can learn from subsequent program executions.
Thus, it would be possible for an attacker to attempt multiple control-flow attacks until the hijack succeeds.
\vspace{-2mm}
\subsection{\fipac Intra Basic Block Protection}

The authors of \fipac describe a mechanism to extend the protection guarantees of \fipac from inter to intra-basic block security~\cite{DBLP:conf/cosade/SchillingNM22}.
By applying a state update after every instruction within a basic block, the subsequently also update the CFI state continuously.
Although this mechanism can be applied around syscalls, it does not add any protection.
With a state update before and after the system call, an attacker can still fault the syscall number or manipulate the \texttt{svc} instruction to perform a \textit{nop} instruction.
Although this attack manipulates the execution of the system call, \fipac's extended intra-basic protection does \textit{not} detect these attacks.
Consequently, it requires a different protection scheme to provide call flow protection for system calls.
\vspace{-4mm}
\section{Design of \fipacos}
\label{sec:fipacos:design}


In this section, we present \fipacos, a mechanism that provides system call flow protection by exploiting a stateful CFI protection scheme.
While \fipacos is generic and compatible with different CFI protection schemes, our design exploits \fipac as the underlying CFI protection scheme.
\Cref{sec:fipacos:compatibility} discusses the compatibility aspects and how \fipacos can be applied to different CFI schemes.
\vspace{-3mm}
\subsection{Requirements for System Call Protection}

The goal of \fipacos is to protect the system call interface to the kernel against software, fault, and combined attacks.
Based on the attack scenario from \Cref{sec:fipacos:threat}, the protection of \fipacos must fulfill the following requirements.

\begin{compactitem}
  \item[\ReqOne]\emph{System Call Number}. Prevent an attacker from manipulating the system call number to a different system call.
  \item[\ReqTwo]\emph{System Call Execution}. Ensure that a syscall cannot be skipped.
  \item[\ReqThree]\emph{System Call Protection}. Ensure the system call dispatcher in the kernel executes the correct system call function.
  \item[\ReqFour]\emph{Dynamic CFI Instrumentation}. Provide a dynamic CFI instrumentation to ensure protection between consecutive program executions.
\end{compactitem}
\vspace{-3mm}
\subsection{System Call Protection}

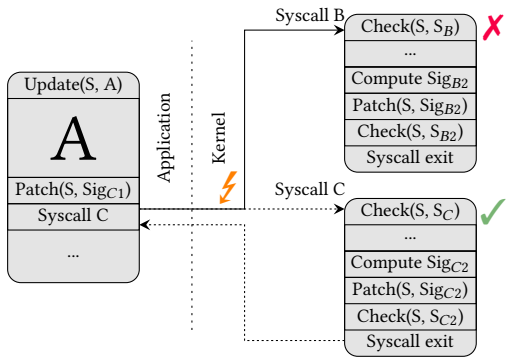
\begin{figure}[t]
  \centering
  \begingroup
  \tikzset{every picture/.style={scale=0.7}}%
  \usetikzlibrary{arrows}
\definecolor{fancygreen}{HTML}{78b473}
\definecolor{fancyred}{HTML}{f70146}
\definecolor{fancygray}{HTML}{a5a5a5}

\newcommand\frameX{0}
\newcommand\frameY{0}
\newcommand\frameLength{12.2}
\newcommand\frameHeight{6}

\newcommand\syscallX{0.1+1.5}
\newcommand\syscallY{4}
\newcommand\syscallLength{2}
\newcommand\syscallHeight{1}

\newcommand\syscallGap{0.3}

\newcommand\AsyscallX{8}
\newcommand\AsyscallY{4}
\newcommand\AsyscallLength{2.5}
\newcommand\AsyscallHeight{3.0}

\newcommand\BsyscallX{8}
\newcommand\BsyscallY{0.5}
\newcommand\BsyscallLength{2.5}
\newcommand\BsyscallHeight{3.0}

\newcommand\userX{0.1+1.5}
\newcommand\userY{1.4}
\newcommand\userLength{2.5}
\newcommand\userHeight{4}

\newcommand\ptrXone{4.1}
\newcommand\ptrYone{2.8}

\newcommand\ptrXtwo{8}
\newcommand\ptrYtwo{0.7}

\begin{tikzpicture}[scale=1.0]
    \tikzset{edge/.style=->, >=stealth, thick};

    \draw[rounded corners=0pt, draw=none] (\frameX-0.05,\frameY+0.5) rectangle node[scale=0.6]{} ++(\frameLength,\frameHeight);
    \draw[rounded corners=5pt, fill=fancygray!30] (\AsyscallX,\AsyscallY-0.5) rectangle node[scale=0.8]{} ++(\AsyscallLength,\AsyscallHeight);
    \draw[rounded corners=0pt, draw=none] (\AsyscallX,\AsyscallY+2) rectangle node[scale=0.8]{Check(S, S$_B$)} ++(\AsyscallLength,\syscallHeight-0.5);
    \draw[rounded corners=0pt] (\AsyscallX,\AsyscallY+1.5) rectangle node[scale=0.8]{...} ++(\AsyscallLength,\syscallHeight-0.5);
    \draw[rounded corners=0pt] (\AsyscallX,\AsyscallY+1.5) rectangle node[scale=0.8]{Compute Sig$_{B2}$} ++(\AsyscallLength,\syscallHeight-1.5);
    \draw[rounded corners=0pt, draw=none] (\AsyscallX,\AsyscallY+1.5) rectangle node[scale=0.8]{Patch(S, Sig$_{B2}$)} ++(\AsyscallLength,\syscallHeight-2.5);
    \draw[rounded corners=0pt] (\AsyscallX,\AsyscallY+0.5) rectangle node[scale=2.8]{} ++(\AsyscallLength,\syscallHeight);
    \draw[rounded corners=0pt] (\AsyscallX,\AsyscallY) rectangle node[scale=2.8]{} ++(\AsyscallLength,\syscallHeight);
    \draw[rounded corners=0pt, draw=none] (\AsyscallX,\AsyscallY) rectangle node[scale=0.8]{Check(S, S$_{B2}$)} ++(\AsyscallLength,\syscallHeight-0.5);
    \draw[rounded corners=0pt, draw=none] (\AsyscallX,\AsyscallY-0.5) rectangle node[scale=0.8]{Syscall exit} ++(\AsyscallLength,\syscallHeight-0.5);
    \draw[rounded corners=5pt, fill=fancygray!30] (\BsyscallX,\BsyscallY-0.5) rectangle node[scale=0.8]{} ++(\BsyscallLength,\BsyscallHeight);
    \draw[rounded corners=0pt, draw=none] (\BsyscallX,\BsyscallY+2) rectangle node[scale=0.8]{Check(S, S$_C$)} ++(\BsyscallLength,\syscallHeight-0.5);
    \draw[rounded corners=0pt] (\BsyscallX,\BsyscallY+1.5) rectangle node[scale=0.8]{...} ++(\BsyscallLength,\syscallHeight-0.5);
    \draw[rounded corners=0pt] (\BsyscallX,\BsyscallY+1.5) rectangle node[scale=0.8]{Compute Sig$_{C2}$} ++(\BsyscallLength,\syscallHeight-1.5);
    \draw[rounded corners=0pt, draw=none] (\BsyscallX,\BsyscallY+1.5) rectangle node[scale=0.8]{Patch(S, Sig$_{C2}$)} ++(\BsyscallLength,\syscallHeight-2.5);
    \draw[rounded corners=0pt] (\BsyscallX,\BsyscallY+0.5) rectangle node[scale=2.8]{} ++(\BsyscallLength,\syscallHeight);
    \draw[rounded corners=0pt] (\BsyscallX,\BsyscallY) rectangle node[scale=2.8]{} ++(\BsyscallLength,\syscallHeight);
    \draw[rounded corners=0pt, draw=none] (\BsyscallX,\BsyscallY) rectangle node[scale=0.8]{Check(S, S$_{C2}$)} ++(\BsyscallLength,\syscallHeight-0.5);
    \draw[rounded corners=0pt, draw=none] (\BsyscallX,\BsyscallY-0.5) rectangle node[scale=0.8]{Syscall exit} ++(\BsyscallLength,\syscallHeight-0.5);
    \draw[rounded corners=5pt, fill=fancygray!30] (\userX,\userY) rectangle node[scale=0.8]{} ++(\userLength,\userHeight);
    \draw[rounded corners=0pt, draw=none] (\userX,\userY+3.5) rectangle node[scale=0.8]{Update(S, A)} ++(\BsyscallLength,\syscallHeight-0.5);
    \draw[rounded corners=0pt] (\userX,\userY+2) rectangle node[scale=2.8]{A} ++(\BsyscallLength,\syscallHeight+0.5);
    \draw[rounded corners=0pt] (\userX,\userY+1.5) rectangle node[scale=0.8]{Patch(S, Sig$_{C1}$)} ++(\AsyscallLength,\syscallHeight-0.5);
    \draw[rounded corners=0pt] (\userX,\userY+1) rectangle node[scale=0.8]{Syscall C} ++(\AsyscallLength,\syscallHeight-0.5);
    \draw[rounded corners=0pt, draw=none] (\userX,\userY+0.5) rectangle node[scale=0.8]{...} ++(\AsyscallLength,\syscallHeight-1.0);
    \draw[edge] (\ptrXone,\ptrYone) -- ++(2,0) -- ++(0,3.4) -- ++(1.9,0);
    \draw[edge, densely dotted] (\ptrXone,\ptrYone) -- ++(3.9,0);
    \draw[edge, densely dotted] (\ptrXtwo,\ptrYtwo-0.4) -- ++(-1.9,0) -- ++(0,2.2) -- ++(-2,0);

    \draw[dotted] (5.1,0.5) -- ++(0,5.5);
    \draw[rounded corners=0pt, draw=none] (\userX+2.5,\userY+1.5) rectangle node[scale=0.8,rotate=90]{Application} ++(1,2.5);
    \draw[rounded corners=0pt, draw=none] (\userX+3.5,\userY+1.5) rectangle node[scale=0.8,rotate=90]{Kernel} ++(1,2.5);

    \draw[rounded corners=0pt, draw=none] (\userX+5,\userY+4.8) rectangle node[scale=0.8]{Syscall B} ++(\BsyscallLength-1,\syscallHeight-0.5);
    \draw[rounded corners=0pt, draw=none] (\userX+5,\userY+1.5) rectangle node[scale=0.8]{Syscall C} ++(\BsyscallLength-1,\syscallHeight-0.5);

    \node[orange] at (5.8,3.2) {\Huge\Lightning};


    \draw[rounded corners=0pt, draw=none] (\AsyscallX+2.5,\AsyscallY+2) rectangle node[scale=1.5,fancyred]{\xmark} ++(\BsyscallLength-1.85,\syscallHeight-0.5);

    \draw[rounded corners=0pt, draw=none] (\AsyscallX+2.5,\AsyscallY-1.5) rectangle node[scale=1.5,fancygreen]{\cmark} ++(\BsyscallLength-1.85,\syscallHeight-0.5);
\end{tikzpicture}%
  \endgroup
  \setlength{\abovecaptionskip}{-1mm}
  \setlength{\belowcaptionskip}{-3mm}
  \caption{System Call protection in \fipacos. Before a syscall, we cryptographically bind the syscall to the CFI state for later verification and second-stage linking in the kernel.}
  \label{fig:fipacos:design:overview}
\end{figure}

To fulfill requirements \ReqOne to \ReqThree, \fipacos introduces a two-step approach cryptographically linking the syscall to the state of the deployed SCFI scheme.
First, at the system call caller site, we cryptographically link the system call origin and which system call we want to execute to the cryptographic CFI state.
Second, at runtime, we perform a second-stage linking operation during the system call operation, confirming that the correct syscall gets executed.
\vspace{-1mm}
\paragraph*{First-Stage System Call Linking.}
We statically identify at compile-time which system call is getting executed for all locations in the program.
To protect the system call, \fipacos binds the syscall to the CFI state, \ie to perform a CFI state update with the system call number.
The system call number is a monotonically increasing number, thus not providing a significant Hamming distance between different system calls.
A single bit-flip on the system call number changes the system call to a different one.
As a result, the system call number cannot safely be used to bind it to the CFI state since faults can easily manipulate the system call to a different one.

To overcome this limitation and perform a safe and secure state update, we need to compute a system call-dependent update value  with a sufficiently large Hamming distance.
In \fipacos, we exploit the cryptographic properties of ARM PA for this purpose.
We use computation of a \pacia operation, with the system call and a random modifier as input, and compute a cryptographic 15-bit patch value for the particular system call.
Due to the cryptographic MAC operations of ARM PA, the patch values for subsequent system call numbers have a large Hamming distance and cannot be computed without having access to the secret ARM PA key.
The computation of those patch values occurs at compile-time or load time and replaces the empty patch values in the binary.

Before executing a system call and jumping to the kernel, we patch the CFI state with the statically computed system call patch, thus performing the \textit{first-stage} linking.
At this point in time, we bind the future execution of the particular system call to the CFI state ahead of executing it.
Performing first-stage linking already provides protection for requirements \textbf{R1} and partly \textbf{R2}.
\vspace{-1mm}
\paragraph*{Second-Stage System Call Linking.}

After linking the system call to the CFI state in the user-space of the program, the system call is executed, and the execution switches into the kernel.
Via dispatching code and the selected system call in the general-purpose register \texttt{w8}, the kernel selects the correct system call function and executes it.
At the end of each system call function, we apply a second patch, \ie the \textit{second-stage} linking to the CFI state, confirming that the previously selected system call was really executed.
This patch value is computed dynamically during the execution of the syscall.
The second linking step ensures that both requirements \textbf{R2} and \textbf{R3} are fulfilled.

In \Cref{fig:fipacos:design:overview}, we summarize \fipacos's system call protection.
A user program performs the first-stage linking and patches the CFI state with a statically computed syscall patch to link the execution of a system call.
The execution transitions to the kernel, which executes the desired system call function.
At the end of the system call, the kernel performs the second-stage linking operation, followed by a CFI check operation.
The later second-stage linking operation only succeeds when the correct system call is linked to the CFI state.
As a result, \fipacos's approach translates system call errors, independent of how they occur, to CFI state errors, which eventually are detected through the checking policy of the selected CFI protection scheme.
Note, \Cref{fig:fipacos:design:overview} includes CFI checks at the beginning and end of the syscall to immediately detect a wrong syscall when entering the kernel and after the syscall's execution.
\vspace{-3mm}
\subsection{Dynamic Instrumentation}

Existing SFCFI protection schemes~\cite{DBLP:conf/eurosp/WernerUSM18,DBLP:journals/compsec/ClercqGUMV17,DBLP:conf/cosade/SchillingNM22,DBLP:conf/host/NasahlSM21} use a static post-processing or encryption phase.
A dedicated post-processing tool recovers the control-flow, computes the patch and check values, and modifies the program.
The static approach with a single encryption key leads to the fact that all executions of the same program use the same CFI values, \eg patches, updates, or checks.
By observing the used CFI-related values, attackers can more easily craft valid CFI states to bypass the control-flow protection.

In \fipacos, we overcome this limitation by splitting up the toolchain and integrating the CFI instrumentation into the kernel.
When starting a program, the ELF loader of the OS identifies a CFI instrumented program.
It generates a random ARM PA encryption key and stores it in the process task structure.
The ELF loader then performs the per-program call unique CFI instrumentation and computes the expected CFI state and all patch values needed to handle the control-flow.
The CFI states are stored along with the process task structure within the kernel.
With this mechanism, subsequent calls to the same program create different encryption keys.
As a result, it guarantees that different CFI values are generated on each new program start, \ie fulfilling requirement \ReqFour.
\vspace{-1mm}
\paragraph*{Kernel Checking Policy.}

In \fipacos, we develop a novel CFI checking policy at the edge of the operating system.
Due to dynamically instrumenting the program when starting it, the operating system exactly knows the expected CFI state for every location of the program.
When a user program now enters the kernel, \eg due to a system call instruction, the kernel, which has access to both the user program state and the expected CFI states, can verify them.
If the current CFI state matches the expected state, the system call continues.
However, if the CFI state deviates from the expected state, a CFI error is detected, and the operating system aborts the program execution.
A CFI check at the end of the syscall confirms the execution of the right syscall.
Apart from system calls, a user program can enter the operating system also via different execution paths.
We include the same checking policy when a timer interrupt is raised, and the kernel is entered.
\vspace{-3mm}
\section{Implementation}
\label{sec:fipacos:implementation}

The prototype implementation of \fipacos consists of two parts.
First, we develop a toolchain to automatically compile and instrument arbitrary C-programs with CFI, including a custom runtime library.
Second, we modify the kernel of the Linux operating system to include the system call verification, the new checking policy, and the dynamic instrumentation on the program start.
\vspace{-3mm}
\subsection{Toolchain}

\vspace{-1mm}
\paragraph*{Compiler.}

We base the toolchain on the modified compiler of \fipac~\cite{fipacllvm}, which is based on the LLVM~\cite{DBLP:conf/cgo/LattnerA04} compiler framework.
We adapt the AArch64 backend of the compiler to instrument the control-flow and embed control-flow meta information in a custom section of the ELF binary.
The compiler inserts the updates for every basic block, inserts patches for control-flow merges, and also deals with call instructions.
Our modified compiler emits a running ELF binary but leaves all patch values for control-flow merges and system calls to be zero.
The necessary post-processing step is shifted to the operating system, which computes all patches at the program start.
Note that the instrumented program does not contain any check instructions as they are part of the transition to the operating system and are performed in the kernel.
\vspace{-1mm}
\paragraph*{C Standard Library.}

System calls are typically invoked via wrapper functions provided by the standard library of the programming language.
This prototype toolchain uses a CFI-instrumented version of the \textit{musl}~\cite{musl} C standard library.
The standard library provides wrapper functions for all system calls or uses system calls directly in different library functions.
We identify every system call in the musl standard library and insert the necessary patch sequence containing an immediate load and the xor-based state update ahead of executing the system call.
\Cref{lst:fipacos:patchedsyscall} summarizes the first-stage linking, where the immediate value for the \texttt{mov} instruction is zero.
When starting the binary, the operating system computes the actual patch value for this system call and fills out the correct load value.

\begin{lstlisting}[float,language=c++,style=aarch64,caption={Patched system call in the musl standard library.},label={lst:fipacos:patchedsyscall}, frame=bt,belowskip=-7mm]
basic_block:
  ...
  mov x0, #...       ; arguments for B system call
  mov x15, #0        ; Zero system call patch
  eor x28, x28, x15   ; Perform a CFI state update
  mov w8, #syscall_B  ; system call number for B
  svc #0             ; Jump to kernel
\end{lstlisting}
\vspace{-3mm}
\subsection{Kernel Support}

\fipacos requires minor modifications to the operating system.
We base the prototype of \fipacos on the Linux kernel in version 5.15.32~\cite{linux-kernel}.
\paragraph*{Dynamic Instrumentation on Program Start.}

On program start, when an instrumented ELF binary is started, \fipacos performs the per-program instrumentation of the program.
First, the kernel generates a random encryption key used for the PA instrumentation.
With the help of control-flow metadata, which is stored along with the ELF binary in a metadata section, we compute the CFI state throughout the program and fill the necessary patch values for justifying signatures.
Furthermore, we compute the syscall- and key-dependent patch values that are used to protect the system call interface.
For every system call in the program, we compute its PAC based on the system call number and user-space program unique modifier.
The resulting PAC value, which is not guessable by the attacker, is filling out the immediate patch value before the syscall.

As discussed, the instrumented program does not contain dedicated CFI check operations as they are performed when entering the kernel.
Instead, we store the expected CFI state for each program location in the task's kernel structure.
To reduce the storage overhead, we use a \texttt{RangeMap}, to only have one entry for a contiguous range of states, where it does not change.

\paragraph*{System Call Verification.}

During the system call, the user program updates the CFI state with a statically computed cryptographic patch value that depends on the system call number.
The verification that the correct system call gets executed happens in the kernel.
After the system call jumps into the kernel, a dispatcher code selects the correct system call function to be executed.
At the end of every system call function in the kernel, we perform the second-stage linking.
Based on the system call number, we dynamically compute a second patch value dependent on the currently executed system call.
In \Cref{lst:fipacos:dynamicsyscallpatch}, we summarize this operation sequence, where we perform the second-stage linking within the kernel.
To retrieve a cryptographically secure patch value, we exploit ARM PA's \pacia instruction, which takes the system call and a modifier as input operands.
Note that the  modifier used for the kernel update of the CFI state is different from the one used for the first-stage linking in the user program.
This property is essential to avoid attackers being able to skip system calls entirely since patching the CFI state twice with the same value would cancel out and has no permanent effect on the CFI state.
We finally apply the computed patch to the CFI state and clear the lower bits from the system call.

\begin{lstlisting}[float,language=c++,style=aarch64,caption={Dynamically computing the system call patch and removing it from the CFI state at the system call end.},label={lst:fipacos:dynamicsyscallpatch}, frame=bt,belowskip=-7mm]
syscall_A:
  ...
  mov x16, #1          ; Load kernel modifier
  pacia x8, x16        ; Compute system call patch
  eor x28, x28, x15    ; Perform 2nd stage linking
  and	x28, x28, #0xffffffff00000000 ; Clear syscall
  ret                  ; number from CFI state
\end{lstlisting}
\vspace{-1mm}
\paragraph*{Checking Policy at the Kernel Boundary.}

Whenever a user program enters the kernel, \fipacos performs a CFI check to validate if the current CFI state still matches the expected state.
We perform CFI checks on two entering points: During a system call and when a timer interrupt is raised.
With the help of the CFI states stored in a \texttt{RangeMap} within the process structure and the knowledge of the program's current program counter, we look up the expected CFI state for the program location.
If the current CFI state, stored in the register \texttt{x28} of the user program state, diverges from the expected state, a CFI error is raised, and \fipacos stops the program execution.
For syscalls, we perform a second CFI check at the end of the syscall function in the kernel to ensure the syscall was really executed.
\vspace{-3mm}
\section{Evaluation}
\label{sec:fipacos:evaluation}

In this section, we first evaluate the security of \fipacos and show how it provides protection and the defined threat model.
Second, we evaluate the functionality and the performance overhead of the prototype implementation.
\vspace{-3mm}
\subsection{Security Evaluation}
\label{sec:fipacos:secevaluation}

We analyze the security guarantees of \fipacos and show how different attacks within the threat model are mitigated.
\vspace{-1mm}
\paragraph*{Control-Flow Hijacks in the User-Space or Kernel.}

\fipacos provides CFI protection for the user-space application based on the selected underlying CFI protection scheme.
The prototype uses \fipac, a basic-block granular CFI scheme, protecting all direct/indirect branches as well as direct/indirect calls.
The protection domain includes the C standard library, which is fully CFI instrumented.
Consequently, an attacker cannot redirect syscalls in the user-space application by redirecting the control-flow to a different wrapper function of the standard library.
Control-flow attacks in the kernel are detected via the kernel internal CFI protection scheme.

\vspace{-1mm}
\paragraph*{Skipping a System Call.}

When skipping a system call instruction, \ie the \texttt{svc} instruction, the first-stage linking already occurred.
Subsequently, the skipped system call misses the second-stage linking from the kernel, which yields a wrong CFI state, which is detectable through the CFI checking policy.
However, if the entire system call instruction sequence is skipped, \ie first-stage patching and the syscall instruction are omitted, the hijack is still detectable.
As both patch operations are missing on the CFI state, the state is wrong again, and a subsequent CFI check, \eg when the program gets scheduled, detects the invalid state.
In both cases, \fipacos transforms the skipped system call into a CFI error, which manifests itself in a wrong CFI state, which is detectable.

\vspace{-1mm}
\paragraph*{Changing a System Call.}

A fault on the register containing the system call number, or a combined attack, in which the attacker controls the register used to execute the system call, redirects the system call to a different one.
\fipacos protects against both attacks.
By applying the first-stage linking to the CFI state, the correct system call is already bound to its future execution.
Manipulating the system call register, \eg due to a fault or software vulnerability, leads to applying the wrong system call patch to the CFI state.
When the system call is executed, the CFI state for that program differs from the expected state, and the CFI check in the kernel detects the problem and aborts the program.

To bypass a system call, the attacker only has a single chance to change the system call number and manipulate the previous system call patch to correct one for this location.
However, the system call patch is protected via the secret ARM PA key, which the attacker cannot access.
Guessing the PAC leads to a probability of \mbox{$p=\frac{1}{2^{15}}=0.0031\,\%$} for getting the correct patch value, where $15$ is the configured PAC size of our prototype implementation.
Furthermore, due to the dynamic instrumentation on the program startup, the system call patches always differ between subsequent calls of the same program.
As a result, the attacker cannot learn new patch information between subsequent program calls.
\vspace{-2mm}
\subsection{Functional Evaluation}
\label{sec:fipacos:functionalevaluation}

To validate the functional correctness of \fipacos, we emulate the execution on the functional simulator QEMU~\cite{qemu} in version 7.0.0.
Since this simulator currently only supports ARM PA from ARMv8.3-A, we extend it to include ARM PA of ARMv8.6-A to support the CFI protection.
The functional evaluation runs the modified Linux kernel from the prototype and can start and execute instrumented programs, where all system calls are protected.
Within the kernel, the functional simulator performs the second-stage linking and a CFI check to verify the execution of the correct syscall.

To verify the functionality of the countermeasure, we emulated skipping a system call and modifying the system call number.
In both cases, \fipacos detects the attack through the next CFI check since the CFI state became invalid and stops the program execution.
\vspace{-2mm}
\subsection{Performance Evaluation}
\label{sec:fipacos:perfevaluation}

At the time of evaluation, there is no publicly available system supporting ARMv8.6-A needed to run \fipac.
However, to conduct the performance evaluation and to measure the performance impact of \fipacos, we emulate the runtime overhead of PA instructions.
Therefore, we base the performance evaluation on a Raspberry Pi 4 Model B~\cite{rpi4} with 8\,GB RAM configured with a fixed CPU frequency of 1.5\,GHz.
The Raspberry Pi contains an  ARM Cortex-A72 CPU based on ARMv8-A but without Pointer Authentication.
To emulate the overhead of PA instructions, we replace them with a PA-analogue instruction sequence, \ie four consecutive XORs.
Related work~\cite{DBLP:conf/uss/LiljestrandNWPE19,DBLP:journals/corr/abs-1905-10242} evaluated this instruction sequence to mimic the timing behavior of a PA instruction.

\vspace{-2mm}
\paragraph*{Microbenchmark.}

\begin{figure}[t]
  \center
  \includegraphics[width=0.7\linewidth]{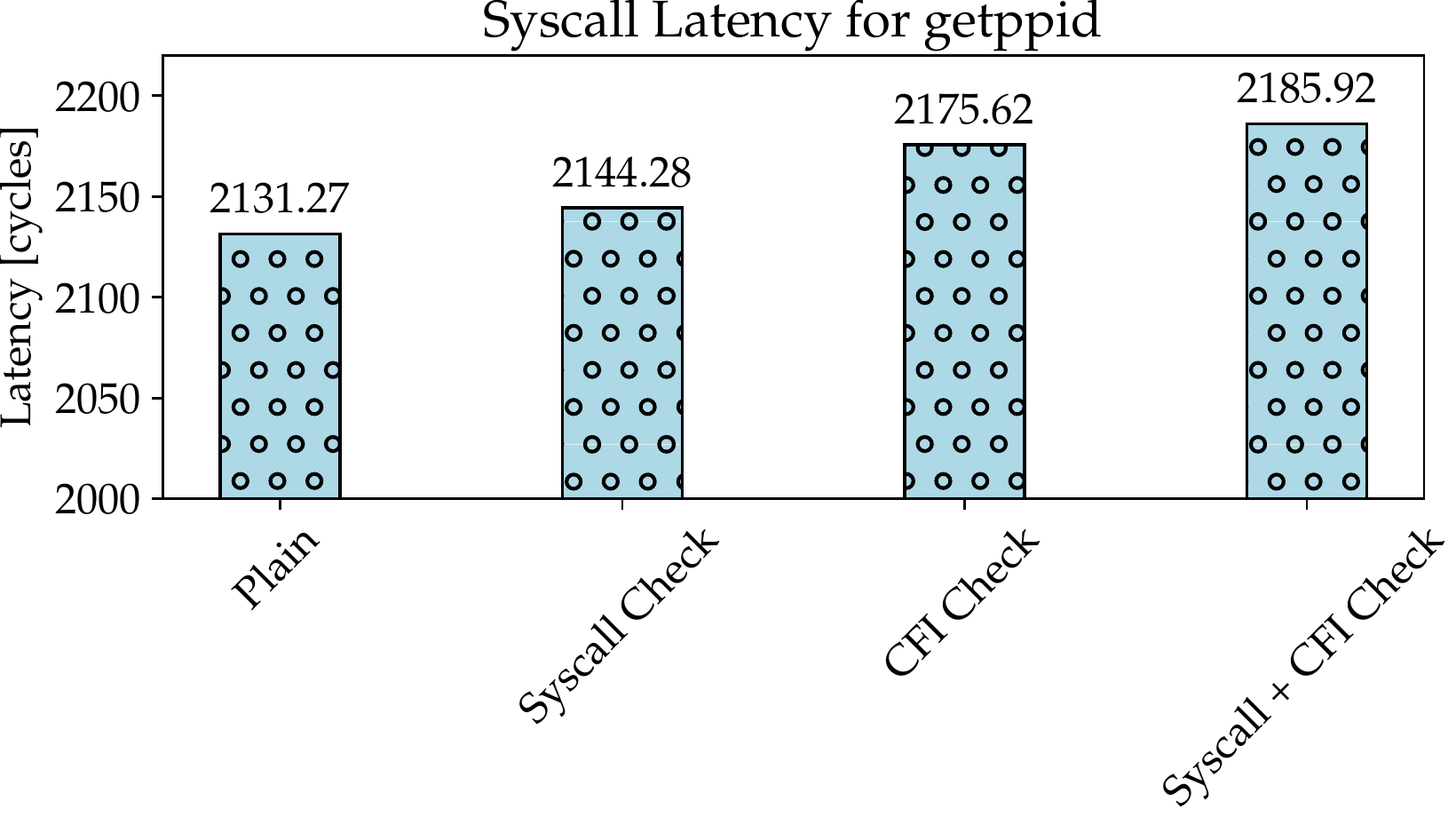}
  \setlength{\abovecaptionskip}{-1mm}
  \setlength{\belowcaptionskip}{-4.5mm}
  \caption{The microbenchmark shows the system call latency of the \texttt{getpid} system call for different kernel configurations. \fipacos increases the system call latency by \microoverhead.}
  \label{fig:fipacos:evaluation:microbenchmark}
 \end{figure}

To evaluate the overhead of \fipacos executing system calls, we perform a simple microbenchmark.
Our benchmark measures the syscall latency of the \texttt{getpid} system call, which is a side-effect-free syscall and is used in related works to benchmark the syscall execution path~\cite{perfbench,DBLP:conf/ccsw-ws/CanellaWG021}.
We execute the system call 10\,million times and measure the system call latency via the processor's inbuilt cycle counter.
\Cref{fig:fipacos:evaluation:microbenchmark} summarizes our evaluation results, showing the syscall latency in different kernel configurations.
On the plain unmodified Linux kernel, we measure an average system call latency of 2131~cycles.
When integrating the system call verification alone, the latency rises to 2144~cycles.
Furthermore, with the CFI checks alone enabled, the latency increases to 2175~cycles.
When both are active, we measure a system call latency of only 2185~cycles, impacting the system call latency by only \microoverhead.

\vspace{-2mm}
\paragraph*{Macrobenchmark.}

\begin{figure}[t]
  \center
  \includegraphics[width=0.9\linewidth]{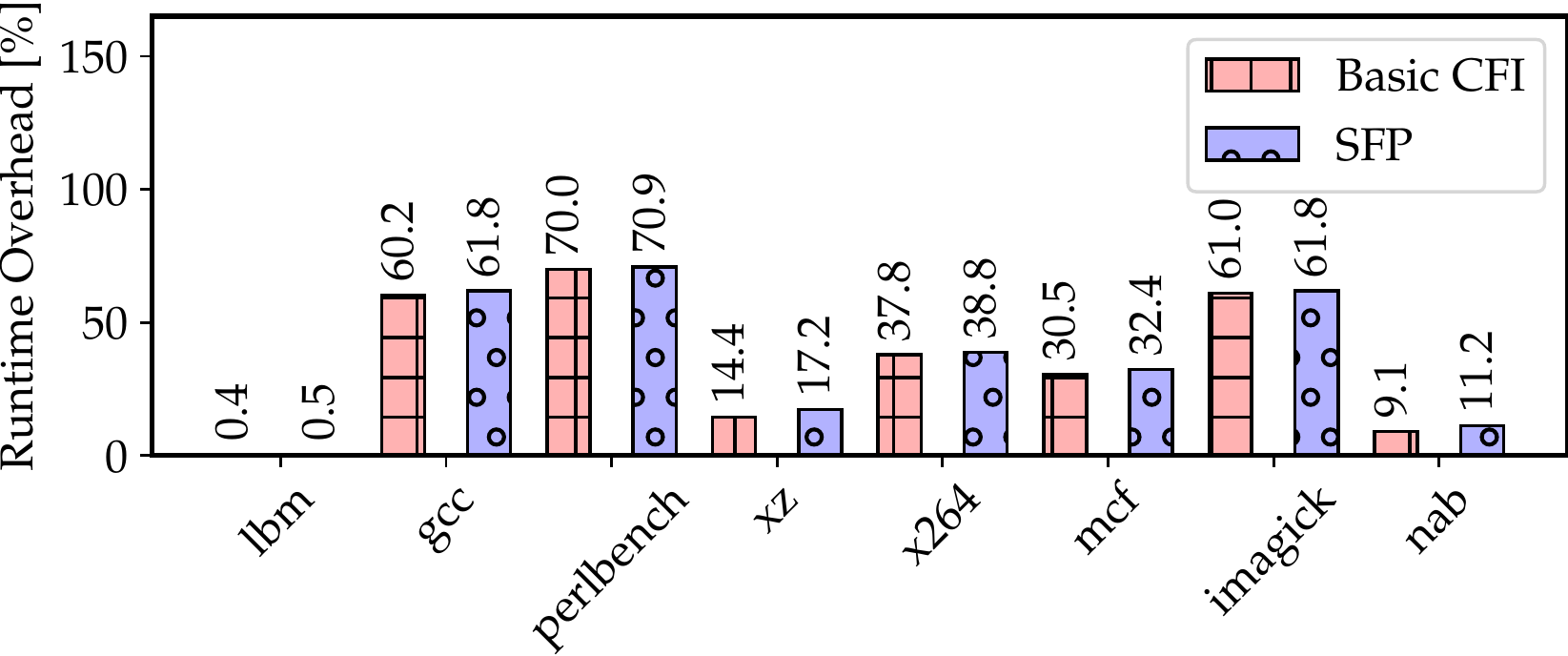}
  \setlength{\abovecaptionskip}{0mm}
  \setlength{\belowcaptionskip}{-4mm}
  \caption{Macrobenchmark shows the performance impact of \fipacos on the SPEC 2017 benchmark. We evaluate the impact of CFI only and \fipacos, including the system call protection.}
  \label{fig:fipacos:evaluation:spec}
 \end{figure}

To demonstrate the applicability of \fipacos on a larger scale, we perform a macrobenchmark on real-world applications.
We compiled the SPECspeed 2017~\cite{Spec2017} benchmark with our toolchain, including only C-based programs.
In \Cref{fig:fipacos:evaluation:spec}, we plot the runtime overheads in two different configurations compared to the plain uninstrumented code.
First, we only include the dynamic verification, including the new CFI checking policy, that verifies the CFI state of user programs when entering the kernel.
Second, we include the syscall protection based on the two-stage linking approach together with the previously evaluated CFI checking policy.

During the evaluation, we measure a geometric mean overhead of \specgeoruntime for the new CFI checking policy and \specgeoruntimefipacos with the system call protection and CFI checking policy in place.
Based on the evaluation of the SPEC 2017 benchmark, we only measure a difference in the overhead of \specdiff between the pure CFI protection and the full system call protection of \fipacos.
This result shows that the dominating part of the overhead comes from the CFI instrumentation, not from the system call protection.
Thus, reducing the overheads of the CFI protection directly influences the performance of \fipacos.

 \vspace{-3.5mm}
\section{Discussion}
\label{sec:fipacos:discussion}

This section discusses prototype limitations and shows how \fipacos is compatible with other CFI protection schemes.
\vspace{-3mm}
\subsection{Dynamic System Call Instrumentation}

In our prototype, we manually instrument all syscalls of the C standard library with the necessary patch instructions, consisting of a load of an immediate patch value followed by applying the patch value to the CFI state.
The immediate value is zero and is set to its concrete value during the dynamic instrumentation of the startup phase of the program.
In a future version of \fipacos, we could instrument the compiler to detect syscall instructions, \ie \texttt{svc}, and then automatically insert the necessary patch sequence.
This enhancement would also include cases where syscalls are invoked manually without the wrapper functions of the standard library.
\vspace{-2mm}

\subsection{CFI Checking Policy Extension}

\fipacos currently performs CFI checks when entering the kernel through a syscall or a timer interrupt.
A future version of this work can extend the CFI checking policy to include all interrupts of the system.
Our microbenchmark shows adding new CFI checks adds minimal overhead to the syscall latency.
Thus, adding additional CFI checks for all interrupt handlers are expected to have minimal impact on the system performance.


\vspace{-2mm}
\subsection{Compatibility}
\label{sec:fipacos:compatibility}

Although \fipacos uses \fipac as the underlying CFI protection scheme, the design or the protection mechanism of \fipacos is generic and compatible with different CFI schemes.
To apply the protection of \fipacos to a different protection scheme, two things are required.
First, the CFI protection scheme must be stateful, and there must be a possibility to manipulate the state, \eg via standard or custom instructions, to inject the system call patch.
Second, it is necessary to  be able to dynamically compute a second system call patch required for the second-stage linking in the kernel.
With these requirements, \fipacos is compatible with existing CFI protection schemes such as SCFP, SOFIA, or any other state-based CFI protection scheme.
\vspace{-2mm}
\section{Related Work}
\label{sec:fipacos:related}

SCFP~\cite{DBLP:conf/eurosp/WernerUSM18} and SOFIA~\cite{DBLP:journals/compsec/ClercqGUMV17} are hardware-assisted control-flow integrity schemes on the instruction level.
They encrypt the program's instruction stream at compile-time, and perform a fine-granular decryption during runtime to retrieve the correct instruction sequence.
In order to deal with the performance penalty, both protection schemes require intrusive hardware changes.
This limits their applicability to small custom embedded processing cores but does not provide protection on a larger scale.

\fipac~\cite{DBLP:conf/cosade/SchillingNM22} is a software-based FCFI protection scheme that exploits the architectural features of recent ARM processors.
This protection scheme instruments all basic blocks of a user program with a running CFI signature, thus providing control-flow integrity at that granularity.
They present three checking policies, \ie where to check whether the running CFI signature still matches the expected one.
However, \fipac only protects the control-flow of the user-space part of the program.
Although \fipac is developed for being used with operating systems, they miss the protection of the system call interface to the kernel.

SFIP~\cite{DBLP:journals/corr/abs-2202-13716} implements coarse-grained syscall flow protection for user-space applications.
They statically identify the possible transitions between different syscalls at compile-time and then enforce that at runtime.
Since SFIP only considers software attackers in their threat model, they fail to protect against fault attacks.
\vspace{-2mm}
\section{Conclusion}
\label{sec:fipacos:conclusion}

In this work, we presented \fipacos, a protection mechanism that provides system call flow protection on top of ordinary CFI, protecting the interface to the kernel against both software and fault attacks.
We show that an already employed CFI protection scheme can be used as a versatile tool to protect the system call interface to the kernel.
Furthermore, we present a new CFI checking policy at the edge of the kernel to verify the CFI state for all transitions to the kernel.
Combined with a dynamic CFI instrumentation on program startup, the attacker cannot learn CFI or system call-related information from subsequent program executions.
We showed a prototype implementation comprising an LLVM-based toolchain to automatically instrument arbitrary programs and protect all system calls.
A modified Linux kernel running on a Raspberry Pi evaluation setup is used to show the applicability of \fipacos to real-world programs.
Our evaluation based on a microbenchmark and on the SPEC 2017 application benchmark shows an average runtime overhead of \specgeoruntimefipacos, which is only an increase of \specdiff compared to plain CFI protection.
This slight increase in the performance impact shows the effectiveness of \fipacos for protecting all system calls of a program.

\ifanonymous
\else
\section*{Acknowledgments}
This work has been supported by the Austrian Research Promotion Agency (FFG) under grant number 888087 (SEIZE).
\fi

\bibliographystyle{ACM-Reference-Format}
\balance
\bibliography{bibliography}

\end{document}